\documentclass[preprint,12pt]{elsarticle}

\usepackage{amssymb}
\usepackage{graphicx,epsfig,rotating,color}

\journal{Chaos, Solitons \& Fractals }

\begin{document}

\begin{frontmatter}





\title{Patterns of trading profiles at the Nordic Stock Exchange. A correlation-based approach.}

\author[label1]{Federico Musciotto}
\author[label1]{Luca Marotta}
\author[label1]{Salvatore Miccich\`e}
\author[label2]{Jyrki Piilo}
\author[label1,label3]{Rosario N. Mantegna}

\address[label1]{Dipartimento di Fisica e Chimica, Universit\'a degli Studi di Palermo, Viale delle Scienze, Ed. 18, I-90128, Palermo, Italy}
\address[label2]{Turku Centre for Quantum Physics, Department of Physics and Astronomy, University of Turku, FI-20014 Turun yliopisto, Finland}
\address[label3]{Center for Network Science and Department of Economics, Central European University, H-1051, Budapest, Hungary}



\begin{abstract}
We investigate the trading behavior of Finnish individual investors trading the stocks selected to compute the OMXH25 index in 2003 by tracking the individual daily investment decisions.  We verify that the set of investors  is a highly heterogeneous system under many aspects. We introduce a correlation based method that is able to detect a hierarchical structure of the trading profiles of heterogeneous individual investors. We verify that the detected hierarchical structure is highly overlapping with the cluster structure obtained with the approach of statistically validated networks when an appropriate threshold of the hierarchical trees is used. We also show that the combination of the correlation based method and of the statistically validated method provides a way to expand the information about the clusters of investors with similar trading profiles in a robust and reliable way.
\end{abstract}


\end{frontmatter}


\section{Introduction}
\label{Int}

The price discovery process in financial markets is the result of the action of many heterogeneous investors \cite{Hommes2006,Samanidou2007,Chakraborti2015}. Understanding how this collective task is achieved is a long standing problem in finance and econophysics. The heterogeneity of investors has been theoretically considered and empirically verified in a series of studies. For example several authors have described investors in stylized categories such as the ones of fundamentalists and chartists \cite{Frankel1990,Chiarella1992,Kirman1993,Lux1995,Brock1998,Lux1999} . In other cases the description has been done in terms of contrarian \cite{Chan1988} and momentum \cite{Chan1996} investors or in terms of informed and uninformed investors (for example see \cite{Grossman1976,Roll1984,Kyle1985,Hasbrouck2007}).

Several studies have considered investment profiles of individual investors. Examples are \cite{Barber2008} that investigated profits of Taiwanese individual investors, \cite{Lachapelle2010} investigating the average transaction value and average portfolio value of individual investors, \cite{Kirilenko2014} that analyzed the detailed trading decisions of single investors acting in the E-mini S\&P 500 stock index futures market during  the Flash Crash of May 6, 2010, \cite{Tumminello2012} that introduced a network based method to characterize specific trading profiles of clusters of investors trading the Nokia stock at the Nordic Stock Exchange, \cite{Feiren2013} that investigated the order splitting of large orders of individual investors, \cite{Challet2013} presenting a robust measure of the contrarian behavior of retail investors, \cite{Bohlin2014} investigating the relation between investors holds and investors profile for Swedish shareholding, and \cite{Lillo2015} where authors study the impact of news on the trading behavior of different categories of individual investors. 

Other papers \cite{Zovko2007,Lillo2008,Vaglica2008,Moro2009,Carollo2012} have investigated the trading behavior of market members of some exchanges such as the Spanish Stock Exchange infrastructure or the London Stock Exchange. Market members are not individual investors but often they show a specialized profile towards some category of investors.

In this paper we investigate the trading behavior of Finnish individual investors with respect to their investment towards the stocks selected to compute the OMXH25 index, i.e. the set of most capitalized and liquid stocks of the Helsinki venue of the Nordic Stock Market. Individual investors are legal entities and can be companies, financial institutions, households, governmental organizations, etc. We are able to track the individual investment decision on a daily basis and therefore our study focuses on a time scale that is a daily time scale or longer. The major observation is that the system is highly heterogeneous under many aspects. For this reason it is important to device data mining methodologies that are robust with respect to the heterogeneity of the different investors and of their trading profiles. In ref. \cite{Tumminello2012} a method based on the concept of statistically validated networks \cite{Tumminello2011} was proposed. Here we introduce another method based on the properties of a dissimilarity measure estimated between the trading profiles of heterogeneous investors. We show that our correlation based method is able to detect a hierarchical structure of the trading profiles of heterogeneous individual investors  and that the detected hierarchical structure is highly overlapping with the cluster structure obtained with the approach of statistically validated networks when an appropriate threshold of the hierarchical trees is used. We also show that the combination of the two methods provides information about the clusters of investors which is wider than the information obtained with the statistically validated network.
 
 The paper is organized as follows. In Section 2. Database we describe the database investigated in this study, and the categorical variables used for the trading activity of the investors. In Section 3.Trading profiles of investors we introduce the correlation based approach used to detect the hierarchical structure of the trading profiles of individual investors.  In Section 4. Over-expressed trading profiles, we apply the statistically validated network approach to the OMXH25 stocks and we present the results obtained. In Section 5. Comparison of clusters, we compare the clusters obtained with the two methods and propose a methodology to combine them to obtain a wider information on the structure of clusters of investors trading in a financial market.  Finally, in Section 6 we briefly present our  conclusions.

\section{Database}
\label{DB}

We have access to a database maintained by the Euroclear Finland (previously Nordic Central Securities Depository Finland) which is the central register of shareholdings for Finnish stocks and financial assets in the Finnish Central Securities Depository (FCSD). The register contains the shareholdings in FCSD stocks of all Finnish investors and of all foreign investors asking to exercise their vote right, both retail and institutional. The database records official ownership of companies and financial assets on a daily basis according to the Finnish Book Entry System. The records include transactions, executed in worldwide stock exchanges and in other venues, which change the ownership of the assets. The database has associated a certain amount of metadata. Specifically, it classifies investors into six main categories: a) non-financial 
corporations, b) financial and insurance corporations, c) general governmental organizations, d) non-profit institutions, e) households, and f) foreign organizations. 
The database is collected since January 1st, 1995. This database has also been investigated by Grinblatt and Keloharju in a series of studies \cite{Grinblatt2000,Grinblatt2009} on the trading characteristics of individual and institutional investors, and on behavioral aspects of individual investors.

In this paper we investigate investment decisions performed during the 2003 calendar year. The database covers all the stocks traded at the Helsinki venue of the Nordic Stock Exchange. Here we investigate the investment decisions concerning 23 of  the 25 stocks composing the OMXH25 market index in 2003\footnote{We originally planned to investigate all the 25 stocks that were used to compute the index but we were not able to find the time history of stock price for two of them.}. 

For legal reasons, the database treats Finnish domestic investors (or foreign investors asking to exercise their vote right) in a different way from foreign investors. In fact, while the database contains very detailed information about the Finnish domestic investors, foreign investors can choose to use nominee registration. In this last case, the investor's book entry account provider, for example a bank, aggregates all the transactions from all of its accounts, and a single nominee register coded identity contains the holdings of several foreign investors. \footnote{If an institution can trade both for itself and also on behalf of nominee registered investors, we split its trading activity in two distinct IDs, one regarding its activity as a Finnish investor and one when it trades for nominee registered investors (labeled as NR).}

In this paper we investigate the trading decisions of investors trading the stocks included in the OMXH25 index during 2003 (a set of 253 daily records). The total number of investors is $105,005$. 
The trading activity of different investors is highly heterogeneous. There are many investors which are acting only a few times during the considered time period. 
Information about the complete set of investors is summarized in Table \ref{Tab1}.

\begin{sidewaystable}
\centering
\caption{Summary statistics of the number of different single investors making at least one transaction for each of the 23 of the OMXH25 stocks during 2003. The different stocks are labeled with their tick symbol. The single investors are grouped in six category: Corporations (non financial corporations), Financial, Financial NR (i.e. those financial institutions that are trading for nominee registered not using their right to vote), Foreign organizations, Governmental organizations, Households, and Non-profit organizations. The last row provides the total number of distinct Finnish investors trading the corresponding stock. The last column provides the total number of investors trading at least one stock of the OMXH25. The total of the column is not the sum of each category row because the same investor can trade different stocks.\newline}
\label{Tab1}
\resizebox{\columnwidth}{!}{      
      \begin{tabular}{lcccccccccccc}
        \hline
          Category & AMEAS	& ELIAV	&ELQAV	&FUM1V	&HUH1V	&KCI1V	&KESBV	&KONBS	&MEO1V	&NDA1V	&NOK1V	&NOR1V\\ 
 Corporations	&624	 &1161	&684	&816	&737 &	339& 	812& 	766& 810&	1082&	3238&	403\\
 Financial&	107& 	121& 	70&	133& 	116& 	82&	95&	102& 	133& 	136& 	212& 	80\\
 Financial NR&	18&	18&	16&	17&	18&	16&	15&	18&	18&	16&	22&	15\\
 Foreign Org.&	39&	56&	36&	68&	57&	21&	32&	62&	49&	97&	434&	19\\
 Governmental&	60&	52&	38&	59&	54&	42&	42&	53&	65&	65&	88&	43\\
 Households	&3439	&14742	&5201	&10372	&5045	&1652	&7345	&6018	&6808	&13374	&37476	&2996\\
Non-profit&	132& 	108& 	55&	182& 	173& 	67&	196& 	207& 	167& 	207& 	259& 	85\\
Total&	4419&	16258&	6100&	 11647&	6200&	2219	&8537&7226&	8050&	14977&	41729&	3641\\
 \hline
        \hline
 Category & OUT1V	&POH1V	&POS1V	&RTRKS	&SAMAS	&STERV	&TIE1V	&TLS1V	&UNR1V	&UPM1V	&WRT1V	&Total\\
	Corporations &1426&	528&	 701&	528& 	924& 	977& 	886& 	1216	&323&	1874	& 672&	6110\\
Financial& 	128& 	58&	110& 	112& 	120& 	118& 	133& 116&	82&	181& 	82&	300\\
Financial NR&	18&	16&	16&	18&	18&	19&	18&	20&	16&	20&	17&	22\\
Foreign Org.&	85&	32&	28&	38&	63&	55&	50&	113&	 34&	164& 41&	772\\
Governmental& 	67&	31&	47&	51&	62&	76&	58&	58&	48&	87&	35&	114\\
Households	&11831	&5843	&5004	&3821	&6600	&8301	&6290	&16003	&1635	&21033	&7285	& 96900\\
Non-profit&	237& 	64&	55&	80&	159& 	152&	 121&	96&	92&	303& 	104& 787\\
Total&	13792&	6572&	5961&	4648&	7946&	9698&	7556&	17622&	2230&	23662&	8236& 105005\\
\hline
\hline
      \end{tabular}
}
\end{sidewaystable}

In the following analyses done with similarity measure, we set a limit to the minimum number of investment decisions to be observed by an investor during 2003 to analyze his trading profile. Our choice is motivated by the need to be able to estimate a similarity measure which is minimizing the discretization role associated with the presence of a very limited number of attributes. Specifically, in the similarity based tests we consider only those investors who have 
traded one of the OMXH25 stocks at least 5 times during 2003. The summary statistics of these investors is given in Table \ref{Tab2}

Since the set of selected investors is still characterized by a very high degree of heterogeneity both in the level of trading activity and in the characteristic of the investor we need to select and use methods of analysis that are robust with respect to the heterogeneity of the attributes of the different investors. This need motivates us to analyze the trading activity of individual investors in terms of categorical variables.

\begin{sidewaystable}
\caption{Summary statistics of the number of different single investors making at least five transactions for each of the 23 of the OMXH25 stocks during 2003.The different stocks are labeled with their tick symbol provided at the top of the column. The single investors are grouped in six category: Corporations (non financial corporations), Financial, Financial NR (i.e. those financial institutions that are trading for nominee registered not using their right to vote), Foreign organizations, Governmental organizations, Households, and Non-profit organizations. The last row provides the total number of distinct Finnish investors trading the corresponding stock. The last column provides the total number of investors trading at least one stock of the OMXH25. The total of the column is not the sum of each category row because the same investor can trade different stocks.\newline}
\label{Tab2}
\resizebox{\columnwidth}{!}{
      \begin{tabular}{lcccccccccccc}
        \hline
          Category & AMEAS	& ELIAV	&ELQAV	&FUM1V	&HUH1V	&KCI1V	&KESBV	&KONBS	&MEO1V	&NDA1V	&NOK1V	&NOR1V\\ 
 Corporations	&44&	83&	98&	65&	67&	29&	58&	81&	105&	144&	1048&	40\\
 Financial&41	&57	&40	&61	&43	&33	&32	&53	&54	&54	&116&	44\\
 Financial NR&	14	&13	&12	&14	&13	&12	&12	&15	&15	&13	&20	&10\\
 Foreign Org.&6	&5	&9	&8	&5	&5	&2	&7	&8	&6	&68	&6\\
 Governmental&11	&21	&8	&23	&12	&13	&9	&11	&25	&31	&60	&15\\
 Households	&71	&194&	391&	214	&110&	44	&173&	227	&442	&468&	6461&	110\\
 Non-profit&11	&12	&13	&9	&10	&7	&6	&14	&17	&17	&51	&13\\
Total&198	&385&	571	&394&	260	&143&	292	&408&	666	&733	&7824	&238\\
 \hline
        \hline
 Category & OUT1V	&POH1V	&POS1V	&RTRKS	&SAMAS	&STERV	&TIE1V	&TLS1V	&UNR1V	&UPM1V	&WRT1V	&Total\\
Corporations &146	&58	&70	&50	&93	&198&	116	&123&	25	&347	&50	& 1310\\
Financial& 51	&24	&47	&49	&54	&65	&55	&46	&29	&80	&32	& 150\\	
Financial NR&	16	&13	&12	&13	&14	&15	&14	&14	&12	&15	&13	& 20\\
Foreign Org.&	6	&5	&6	&5	&7	&11	&8	&13	&4	&14	&6	& 85\\
Governmental&31	&9	&12	&21	&27	&34	&18	&23	&7	&43	&7	& 69\\
Households &416	&186&	352	&91	&340&	781	&454&	671	&23	&1271	&167& 8320\\
Non-profit&20	&8	&14	&10	&18	&24	&14	&16	&5	&29	&4	& 78\\
Total&686	&303&	513	&239&	553	&1128	&679&	906	&105&	1799&	279	& 10032\\	
\hline
\hline
      \end{tabular}
}
\end{sidewaystable}

\subsection{Nominal variables characterizing the trading activity}\label{categorical}

Since we aim to compare the trading profile characterizing each active investor, we use discrete nominal variables describing its trading activity.  Specifically, we use the nominal variables introduced in Ref. \cite{Tumminello2012}. These variables are defined as follows: for each investor $i$, each stock $k$, and each trading day $t$, we consider the volume sold $V_s(i,k,t)$ and the volume purchased $V_b(i,k,t)$ of stock $k$ by the investor at day $t$. For each stock and for each day, this information is then converted into a nominal variable with 3 states: primarily buying \emph{b}, primarily selling \emph{s}, buying and selling \emph{bs} such that all positions will usually be essentially closed before the market close.  The nominal variables are obtained by considering the ratio
\begin{equation}
r(i,k,t)=\frac{V_b(i,k,t)-V_s(i,k,t)}{V_b(i,k,t)+V_s(i,k,t)}.
\end{equation}
For each stock $k$, we assign an investor a primarily buying state \emph{b} when $r(i,k,t)>\theta$,  a primarily selling state \emph{s} when $r(i,k,t)<-\theta$, and a buying and selling state \emph{bs} when $ -\theta \le r(i,k,t) \le \theta$  with $V_b(i,k,t)>0$ and $V_s(i,k,t)>0$. 
In this work we use $\theta=0.25$ for the threshold value.

\section{Trading profiles of investors}
\label{SXX}
The first analysis we perform concerns the similarity of different pairs of investors in the trading activity and in the trading profile. For this type of investigations we consider the set of investors that have done at least five transactions in the considered stock of OMXH25 in 2003. A summary statistics of these investors is given in Table \ref{Tab2}. 
\subsection{Trading profile clustering by correlation}
We evaluate the degree of similarity in the trading profile of investors by constructing for each selected investor and for each stock of the OMXH25 a vector of trading actions. This vector will have a number of components equal to three times the number of trading days of 2003 (which is 253x3=756). The first 253 components will carry the information whether the investor was buying (b state) on a specific day,  the second 253 components will carry the information whether the investor was selling (s state) on a specific day, and the third 253 components will carry the information whether the investor was buying, selling and closing the position (bs state) on a specific day. The final vector is therefore a binary vector of 1s and 0s. To investigate these vectors we need a similarity measure which is robust with respect to the asymmetric presence of the two attributes (1s are rarely present whereas 0s are highly observed in most of the cases). We therefore use the Jaccard coefficient as similarity measure because this measure it is known to be robust for asymmetric binary vectors \cite{Han2011}.

\begin{center}
\begin{figure}
\includegraphics[scale=0.45]{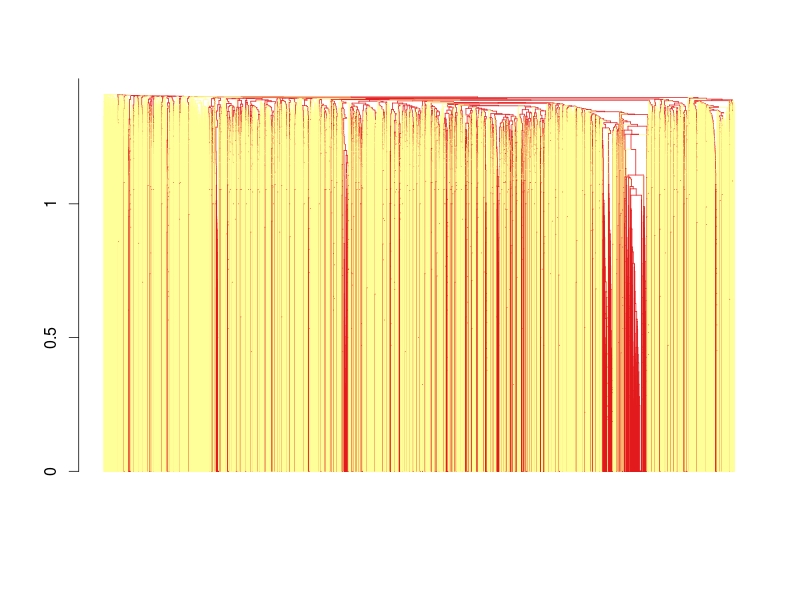}
\caption{\scriptsize{Average linkage hierarchical tree of the trading profile similarity of investors trading Nokia in 2003. The selected investors have performed at least 5 transactions of the Nokia stock during 2003. The presence of large clusters comprising hundreds of investors is clearly detected (see the region of the hierarchical tree highlighted in Red color).The investors labeled with red lines belong to clusters obtained by setting a cutting threshold equal to $d=1.09$ (see Sect. \ref{COC} for details).}}
\label{F1}
\end{figure}
\end{center}

\begin{center}
\begin{figure}
\includegraphics[scale=0.45]{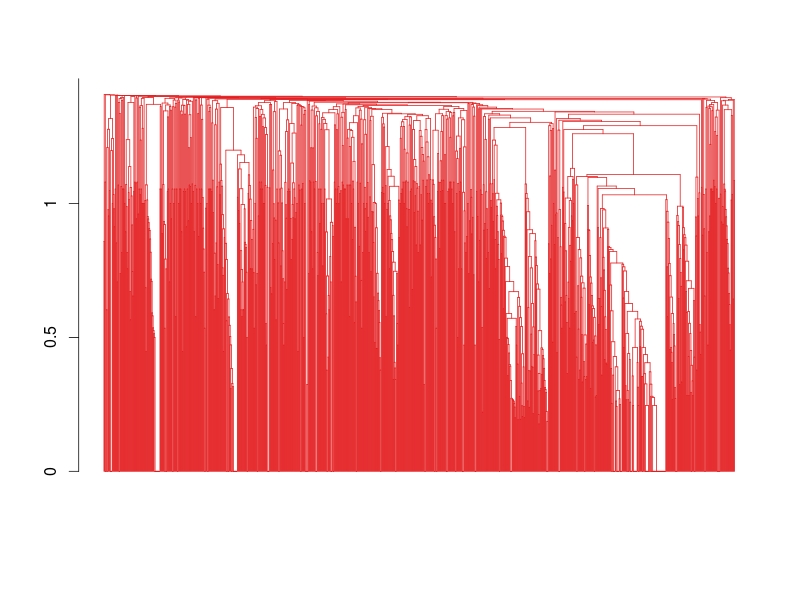}
\caption{\scriptsize{Regions of the average linkage hierarchical tree of Fig. \ref{F1} that are selected by using the threshold $d=1.09$ (see Sect. \ref{COC} for details). All remaining investors are removed from the present tree. The two big clusters seen in the right part of the figures are the two clusters highlighted in red in the right part of Fig. \ref{F1}.}}
\label{F2}
\end{figure}
\end{center}

In Fig. \ref{F1} we show the hierarchical tree obtained with the average linkage hierarchical clustering with a dissimilarity measure defined as $d_{i,j}=\sqrt{2(1-J_{i,j})}$ where $J_{i,j}$ is the Jaccard coefficient observed between investor $i$ and investor $j$.
Nokia has 7824 investors trading at least five times during 2003 and therefore the complete visualization of a hierarchical tree of 7824 elements is not simple in a limited space. In fact the details of the hierarchical tree can be appreciated only by zooming the part of interest. In Fig. \ref{F2} we show a region of the hierarchical tree involving 1419 investors. The hierarchical structure of the similarity of investment profile can be appreciated at this level of details.

A behavior similar to the one observed for the stock of Nokia it is observed for all the other 22 investigated stocks. The use of categorical variables together with the choice of the Jaccard coefficient allow us to detect the hierarchical clustering structure of the trading profile of the different investors. 
In addition to the average linkage clustering algorithm other popular hierarchical clustering algorithms are the average single and the complete linkage clustering algorithms. 

\section{Over-expressed trading profiles}
\label{OEP}
In a previous study \cite{Tumminello2012} a different approach based on statistically validated networks \cite{Tumminello2011} was used to detect clusters of investors characterized by the same trading profile when trading the Nokia stock over a time period longer than 5 years. With the methodology proposed in  \cite{Tumminello2012} authors were able to detect clusters of investors characterized by similar trading profile. In the present study we wish to compare the clusters obtained by using the approach of Ref.  \cite{Tumminello2012} with the information that can be  obtained by the hierarchical structure which is detected starting from the dissimilarity measure we have defined in the previous Section.

The method of the statistical validation of the co-occurrence of categorical variables is the same as in Ref. \cite{Tumminello2012} and works as follows. For each pair $i$ and $j$ of investors we focus our attention on the intersection of the corresponding activity periods. We call $N_T$ the length of this intersection. In the intersection period of traders' activity, let us call $N_A$ ($N_B$) the number of days when investor $i$ ($j$) is in the state $A$ ($B$) and indicate by $N_{A,B}$ the number of days when we detect co-occurrence of state $A$ for investor $i$ and state $B$ for investor $j$. Under the null hypothesis of random co-occurrence, the probability of observing $X$ co-occurrences of $A,B$ is described by the hypergeometric distribution, $H(X|N_T,N_A,N_B)$ \cite{Tumminello2011}. By knowing the probability of $X$ co-occurrences, it is possible to estimate a p-value for each pair of investors and for each combination of the investigated states $A$ and $B$. 
Specifically, for each pair of states $A$ and $B$, the p-value is
\begin{equation}
\label{pval}
p(N_{A,B})=1-\sum_{X=0}^{N_{A,B}-1}H(X|N_T,N_A,N_B).
\end{equation}
The nine possible combinations of the three trading states between investor $i$ and $j$ are ($i_{b}$,$j_{b}$), ($i_{b}$,$j_{s}$), ($i_{b}$,$j_{bs}$),
($i_{s}$,$j_{b}$), ($i_{s}$,$j_{s}$), ($i_{s}$,$j_{bs}$),
($i_{bs}$,$j_{b}$), ($i_{bs}$,$j_{s}$), 
and ($i_{bs}$,$j_{bs}$).

The statistical validation of the co-occurrence $N_{A,B}$ requires a multiple hypothesis test correction \cite{Miller1981}.
Widely used multiple hypothesis test corrections are the Bonferroni and the control of the False Discovery Rate (FDR) methods. For each stock $s$ we use as Bonferroni correction  $\alpha_b=2~0.01/9(N_s (N_s-1))$, where $N_s$ is the number of investors trading stock $s$ at least once in 2003. A less stringent correction is the one obtained with the control of the FDR \cite{Benjamini1995}, which is calculated as follows: p-values from all the
different k tests  ($k= 9 N_i (N_i-1)$ in the present case) are first
arranged in increasing order ($p_1<p_2<...<p_k$), and then statistical threshold is obtained by finding the largest $k_{max}$ such that $p_{k_{max}}<k_{max} ~ p_b$.
\begin{sidewaystable}
\caption{Summary statistics of the number of different single investors having at least one statistical validation of co-occurrence for each of the 23 of the OMXH25 stocks during 2003.The multiple hypothesis test correction is the Bonferroni correction. The label of stocks and investors' categories is as in Table \ref{Tab1}. The last row provides the total number of distinct Finnish investors having co-occurrences of some daily trading status ($b$, $s$, and $bs$). The last column provides the total number of investors having co-occurrences in at least one stock of the OMXH25. The total of the column is not the sum of each category row because the same investor can have co-occurrences for different stocks.\newline}
\label{Tab3}
\resizebox{\columnwidth}{!}{
      \begin{tabular}{lcccccccccccc}
        \hline
          Category & AMEAS	& ELIAV	&ELQAV	&FUM1V	&HUH1V	&KCI1V	&KESBV	&KONBS	&MEO1V	&NDA1V	&NOK1V	&NOR1V\\ 
 Corporations	&0	&1	&3	&2	&0	&0	&0	&1	&1	&1	&142&	3\\
 Financial&5	&2	&3	&8	&4	&0	&2	&3	&9	&5	&18	&3\\
 Financial NR&1	&0	&1	&1	&0	&0	&0	&1	&1	&2	&1	&0\\	
 Foreign Org.&1	&0	&3	&1	&0	&0	&0	&1	&1	&1	&5	&0\\
 Governmental&0	&1	&0	&1	&0	&0	&0	&0	&4	&1	&16	&8\\
 Households&2&0	&0	&2	&0	&0	&0	&6	&0	&2	&378&	1\\
 Non-profit&0	&0	&0	&0	&0	&0	&0	&0	&0	&0	&16	&2\\
 Total&9	&4	&10	&15	&4	&0	&2	&12	&16	&12	&576&	17\\
 \hline
        \hline
 Category & OUT1V	&POH1V	&POS1V	&RTRKS	&SAMAS	&STERV	&TIE1V	&TLS1V	&UNR1V	&UPM1V	&WRT1V	&Total\\
Corporations&2	&0	&0	&2	&1	&8	&1	&1	&1	&11	&0	&154\\
Financial&4	&0	&2	&5	&4	&12	&3	&4	&1	&10	&0	&31\\ 
Financial NR&0	&1	&0	&0	&0	&0	&0	&1	&0	&0	&0	&3\\
Foreign Org.&0	&1	&0	&0	&0	&0	&0	&1	&0	&0	&0	&6\\
Governmental&4	&3	&0	&6	&2	&12	&0	&1	&0	&10	&0	&18\\
Households&2	&2	&0	&0	&2	&6	&3	&0	&0	&4	&2	&389\\
Non-profit&3	&3	&0	&1	&0	&5	&0	&2	&0	&4	&0	&16\\
Total&15	&10	&2	&14	&9	&43	&7	&10	&2	&39	&2	&617\\
\hline
\hline
      \end{tabular}
}
\end{sidewaystable}

In Table \ref{Tab3} we report the number of investors characterized by at least one co-occurrence of the set of nine investigated for each investigated stock. The multiple hypothesis test correction used in the statistical validation is the Bonferroni correction. The largest number of investors characterized by co-occurrence in the trading profile is detected for the Nokia stock. This is not surprising because the Nokia stock was in 2003 the most traded (as it can be verified in Table \ref{Tab1}) and liquid stock of the OMXH25 index.  The results can also depend on the power of the statistical test that becomes stronger when the number of states to be statistically validated increases.

\begin{figure}
\centering \includegraphics[scale=0.3]{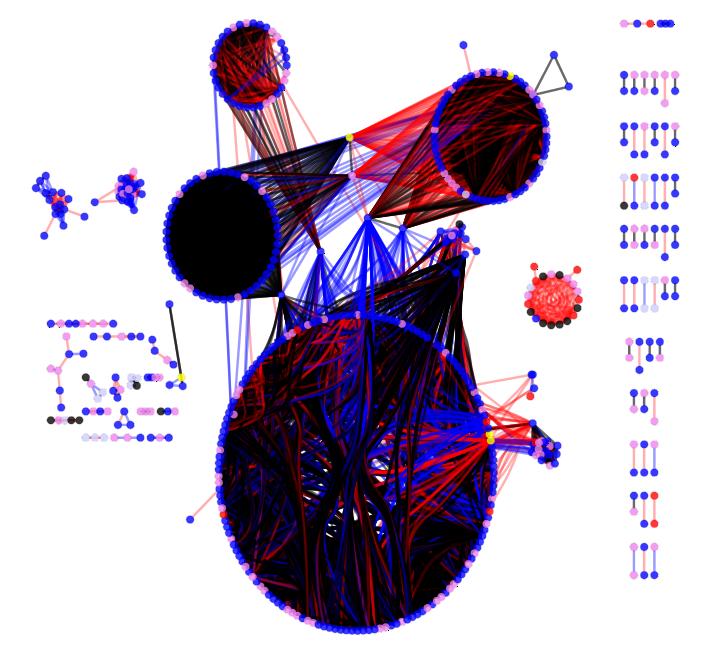}
\caption{\scriptsize{Network of 576 individual investors having statistically validated co-occurrences of trading decisions about the Nokia stock in 2003. The multiple hypothesis test correction is the Bonferroni correction. The category of investor is labeled as follows: Corporations (Violet), Financial (Grey), Foreign organizations (Yellow), Governmental (Black), Households (Blue) and Non-profit (Red). The type of statistically validated co-occurrence is labeled as follows: $i_{b}$,$j_{b}$ (1075 Blue links), $i_{s}$,$j_{s}$ (2468 Red links),  and the simultaneous validation of $i_{b}$,$j_{b}$ and $i_{s}$,$j_{s}$ (10353 Black links).}}
\label{F3}
\end{figure}

In Table \ref{Tab4} we report the results for the multiple hypothesis test correction based on the FDR correction. As expected the number of investors showing statistically validated co-occurrences is increasing and the accuracy of the test is improved although the level of precision might slightly decrease. The number of investors with statistically validated co-occurrence of Table \ref{Tab3} are of course always included in the corresponding entry of Table \ref{Tab4}. 
\begin{sidewaystable}
\caption{Summary statistics of the number of different single investors having at least one statistical validation of co-occurrence for each of the 23 of the OMXH25 stocks during 2003.The multiple hypothesis test correction is the FDR  correction. The label of stocks and investors' categories is as in Table \ref{Tab1}. The last row provides the total number of distinct Finnish investors having co-occurrences of some daily trading status ($b$, $s$, and $bs$). The last column provides the total number of investors having co-occurrences in at least one stock of the OMXH25. The total of the column is not the sum of each category row because the same investor can have co-occurrences for different stocks.\newline}
\label{Tab4}
\resizebox{\columnwidth}{!}{
      \begin{tabular}{lcccccccccccc}
        \hline
          Category & AMEAS	& ELIAV	&ELQAV	&FUM1V	&HUH1V	&KCI1V	&KESBV	&KONBS	&MEO1V	&NDA1V	&NOK1V	&NOR1V\\ 
 Corporations&0	&1	&3	&3	&2	&0	&0	&1	&2	&1	&298&	4\\
  Financial&5	&4	&3	&8	&5	&0	&2	&3	&10	&5	&40	&6\\
  Financial NR&1	&0	&1	&1	&1	&0	&0	&1	&1	&2	&2	&1\\
  Foreign Org.&1	&0	&3	&1	&1	&0	&0	&1	&1	&1	&9	&1\\
  Governmental&0	&1	&0	&2	&0	&0	&0	&0	&5	&3	&39	&9\\
  Households&2	&0	&2	&4	&0	&0	&0	&6	&0	&5	&1098	&3\\
  Non-profit&0	&0	&0	&0	&0	&0	&0	&0	&0	&0	&32	&5\\
  Total&9	&6	&12	&19	&9	&0	&2	&12	&19	&17	&1518	&29\\
 \hline
        \hline
 Category & OUT1V	&POH1V	&POS1V	&RTRKS	&SAMAS	&STERV	&TIE1V	&TLS1V	&UNR1V	&UPM1V	&WRT1V	&Total\\
Corporations&5&	0	&0	&5	&3	&11	&3	&5	&1	&14	&0	&310\\
Financial&7	&0	&2	&7	&5	&15	&3	&4	&1	&13	&0	&48\\
Financial NR&0	&1	&0	&0	&0	&0	&0	&1	&0	&0	&0	&4\\
Foreign Org.&2	&1	&0	&0	&0	&0	&0	&1	&0	&3	&0	&10\\
Governmental&8	&5	&0	&7	&6	&17	&2	&1	&0	&12	&0	&41\\
Households&13	&2	&0	&0	&2	&16	&6	&1	&0	&13	&2	&1116\\
Non-profit&5	&3	&0	&3	&0	&8	&0	&3	&0	&6	&0	&34\\
Total&40	&12	&2	&22	&16	&67	&14	&16	&2	&61	&2	&1563\\
\hline
\hline
      \end{tabular}
}
\end{sidewaystable}  
In Fig. \ref{F3} we show the statistically validated network of Nokia investors obtained with the Bonferroni correction. The network consists of 576 investors. The majority of them are Households but also investors of the other categories are present. Specifically, in the network there are 142 Corporations (Violet node), 18 Financial and 1 Financial NR (Grey node), 5 Foreign organizations (Yellow node), 16 Governmental (Black node), 378 Households (Blue node) and 16 Non-profit (Red node) investors. Direct inspection of the network shows that there there is a large connected component of 346 investors and that all validations of co-occurrences concern  $i_{b}$,$j_{b}$ (Blue link), $i_{s}$,$j_{s}$ (Red link), and the simultaneous validation of $i_{b}$,$j_{b}$ and $i_{s}$,$j_{s}$ (Black link).

As summarized in Tables \ref{Tab3} and \ref{Tab4} the largest detected statistically validated networks are the ones of investors trading the Nokia stocks. For this stock the Bonferroni (i.e. the statistically validated networks obtained with the Bonferroni correction) and FDR (i.e. the statistically validated networks obtained with the Benjamini-Hochberg correction) networks have 576 and 1518 nodes respectively. In the case of the other OMXH25 stocks the Bonferroni and FDR networks are always populated but with a much smaller number of nodes. In fact the second largest statistically validated network is the one of Stora Enso comprising 67 investors. The network was obtained with the FDR correction (see Table \ref{Tab4}).

\begin{figure}
\centering \includegraphics[trim=0cm 7cm 0cm 1cm,  scale=0.40]{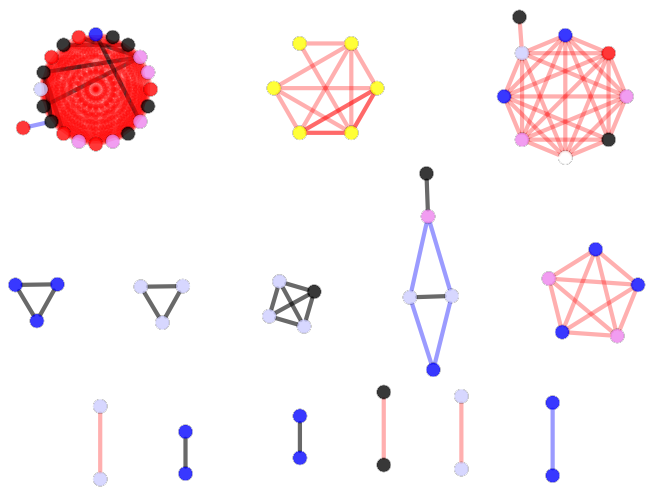}
\caption{\scriptsize{Network of 67 individual investors having statistically validated co-occurrences of trading decisions about the Stora Enso stock in 2003. The multiple hypothesis test correction is the FDR correction.  Links are coded as follows: $i_{b}$,$j_{b}$ (6 Blue links), $i_{s}$,$j_{s}$ (210 Red links),  and the simultaneous validation of $i_{b}$,$j_{b}$ and $i_{s}$,$j_{s}$ (26 Black links)}}
\label{F4}
\end{figure}

This last statistically validated network is shown in Fig. \ref{F4}. In this case the majority of them are Governmental organizations. The investors of different categories are as follows: 11 Corporations (Violet node), 15 Financial (Grey node), 17 Governmental (Black node), 16 Households (Blue node) and 8 Non-profit (Red node) investors. As in the case of the previous example, most of the validations of co-occurrences concern  $i_{b}$,$j_{b}$ (Blue link), $i_{s}$,$j_{s}$ (Red link), and the simultaneous validation of $i_{b}$,$j_{b}$ and $i_{s}$,$j_{s}$ (Black link).  The network does not present a large connected component covering the majority of nodes in fact the largest connected component has only 21 nodes.

Most of the remaining OMXH25 stocks have statistically validated networks of the type observed for Stora stock, i.e. a network a small disjoint clusters.
%
%
%
As in Ref. \cite{Tumminello2012} to obtain the clusters of investors with similar trading profile we apply to all obtained statistically validated networks a widely used community detection algorithm. Specifically we apply the Infomap algorithm \cite{Rosvall2007} to the weighted version of our network where the weight of each link is the number of co-occurrences validated between the two investors (for example in the case when we validate  $i_{b}$,$j_{b}$ and $i_{s}$,$j_{s}$ the weight of the link is set to two). In most cases the clusters observed in the networks are not further partitioned by the algorithm. However, in the case of presence of a highly populated large connected component (as for Nokia) the large component is partitioned by the algorithm.

\begin{center}
\begin{figure}
{\includegraphics[scale=0.32]{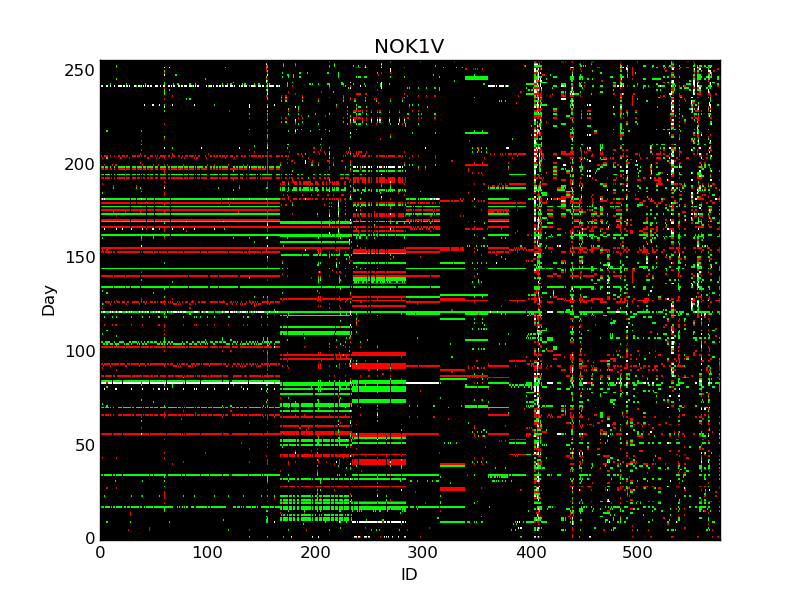}
\includegraphics[scale=0.32]{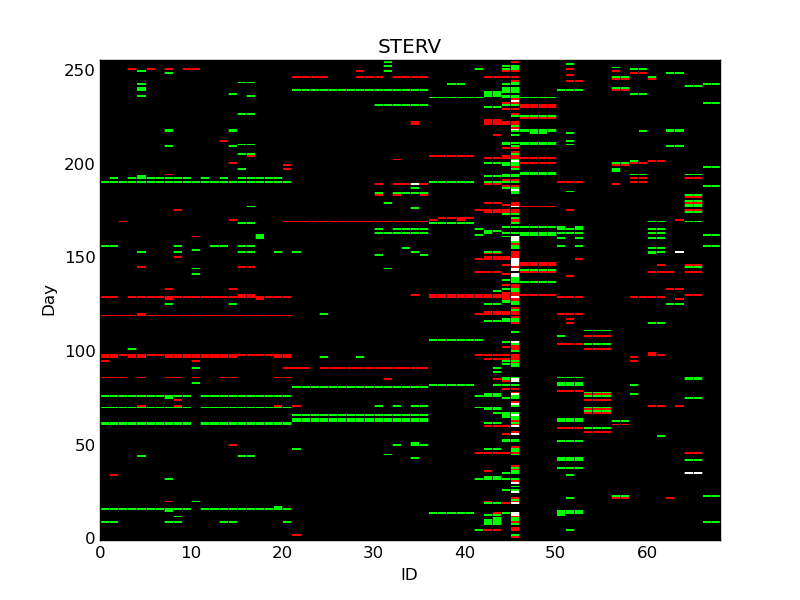}}
\caption{\scriptsize{Color code representation of the trading profile of investors trading the Nokia (left) and the Stora (right) stocks. In the horizontal axis we order different investors whereas the vertical axis is time (in number of trading days). The left panel shows trading profiles of the 576 Nokia investors whose trading co-occurrences were statistically validated with the Bonferroni correction whereas the right panel shows to the 67 Stora investors validated with the FDR correction. A red spot indicates a buy action, a green spot a sell action and a white spot a buy/sell action. Black spots indicate absence of trading. Different trading profiles are easily detectable especially for large clusters.
}}
\label{F5}
\end{figure}
\end{center}
As done in Ref. \cite{Tumminello2011}, we use the partitions obtained to visualize the trading patterns of investors belonging to the detected clusters. In Fig. \ref{F5} we show  the trading profile of Nokia and Stora investors associated with the statistically validated networks shown in Fig. \ref{F3} and \ref{F4}.  In the figure  a red spot indicates a buy action, a green spot a sell action, and a white spot a buy/sell action. A black spot indicates absence of trading for the specific investor and trading day. Fig. \ref{F5} clearly shows the presence of different trading profiles within the clusters of the statistically validated networks that have been in some cases partitioned by the Infomap community detection algorithm.

\section{Comparison of clusters}
\label{COC}
In the previous sections we have shown that both a correlation analysis and a statistical validation procedure are providing information about (i) the clusters of trading profile of different investors (in the case of the statistical validation approach jointly used with the community detection procedure), and (ii) about the hierarchical clustering of the trading profiles. In this section we investigate whether the two types of information overlaps or are rather providing complementary types of information and in case of overlap, we aim to estimate the degree of this overlap.

One difficulty in the comparison of  the two sets of information is that whereas a partition is provided in the case of statistically validated networks, the hierarchical clustering just provides a hierarchical structure that needs to be processed to obtain a partition. A basic way to obtain a partition starting from a hierarchical tree is to cut the hierarchical tree at a given value of the dissimilarity (or similarity) measure and consider the clusters of elements that are connected at distances lower that the cutting threshold. The method is simple and effective but the drawback is that there is no known optimal way to select the threshold level. 

For this reason we decide to apply the following protocol to detect the degree of maximal overlapping between the partition of the SVN and partitions obtained from hierarchical trees. Hereafter we illustrate our portfolio for the set of investors trading the Nokia stock. First we select a hierarchical clustering procedure and obtain the hierarchical tree of the considered system (the 7824 investors investigated with the correlation approach and trading the Nokia stock in 2003). We use as hierarchical clustering procedure the single linkage, the average linkage, and the complete linkage \cite{Han2011}. For each hierarchical tree obtained we set a threshold value and obtain from the tree a partition for the the investors with clusters characterized by a set of distances among the investors which is less then the chosen threshold. The investors linked at higher distances are considered as isolated nodes. From the obtained hierarchical tree partition we select the partition referring to the 576 investors that are also present in the Bonferroni network. We therefore compare the two partitions of 576 investors (one obtained from SVN plus community detection and one obtained by thresholding the hierarchical tree) by computing the Adjusted Rand Index \cite{Hubert1985}. The adjusted Rand Index is an indicator which is widely used for the comparison of partitions. It is a normalized indicator and assumes a value equal to one when the two partitions are identical and a value close to zero when the two partitions are randomly assigned.  
\begin{center}
\begin{figure}
\includegraphics[scale=0.32]{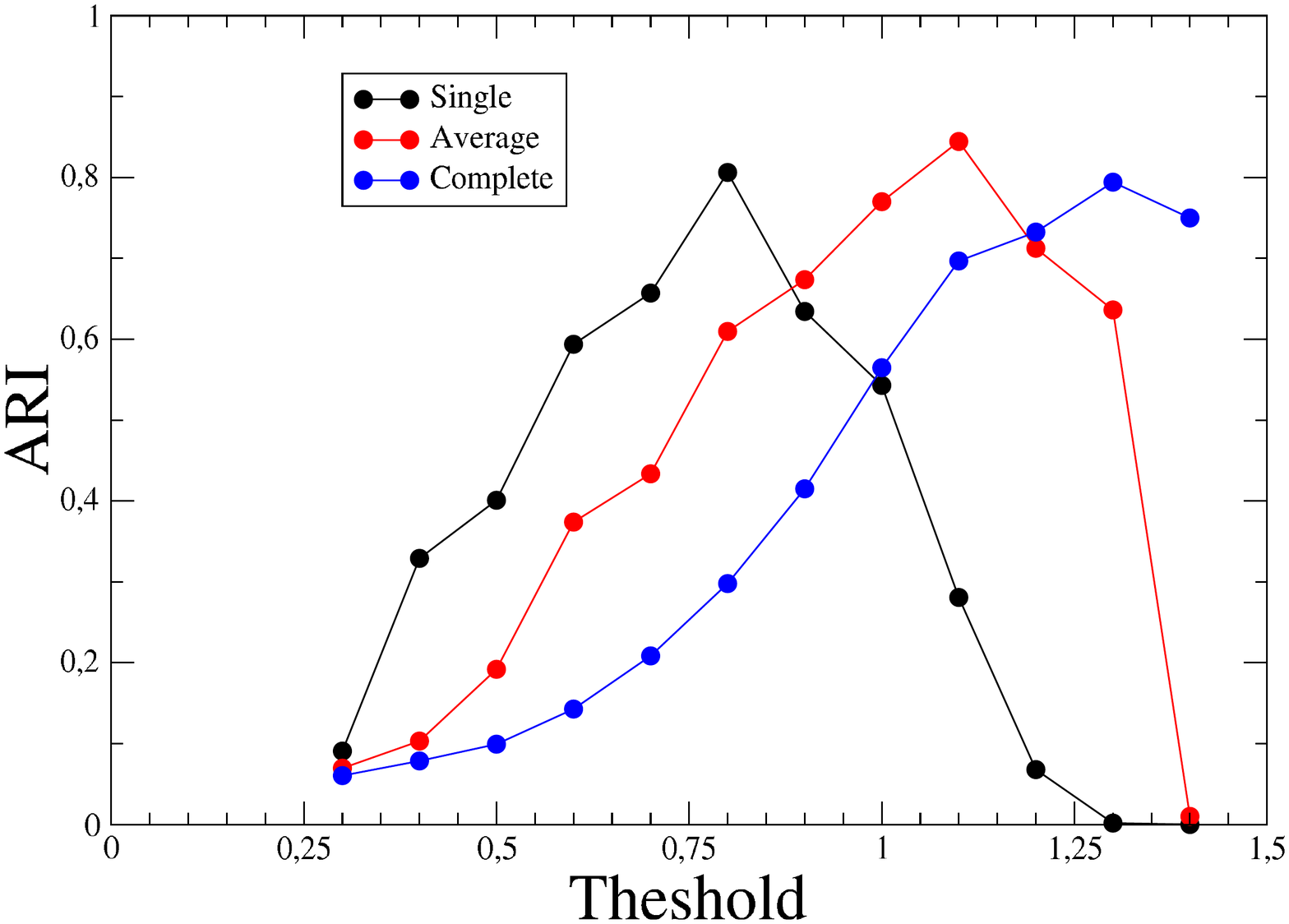}
\caption{\scriptsize{Adjusted Rand Index (vertical axis) between the partition of Nokia investors obtained from the Bonferroni statistically validated network partitioned by the Infomap algorithm with the partition of the same investors obtained by performing hierarchical clustering algorithms on the dissimilarity measure of Nokia investors and selecting different values of the cutting threshold (horizontal axis). The comparison is performed for the Single linkage (black circles), average linkage (red circles), and complete linkage (blue circles). Lines are drawn to guide the line.
}}
\label{F6}
\end{figure}
\end{center}
In Fig. \ref{F6} we show the Adjusted Rand Index between the partition of the Bonferroni statistically validated network partitioned by the Infomap algorithm with the partition of the same investors obtained  from hierarchical clustering algorithms for different values of the cutting threshold (horizontal axis). The comparison is performed for the Single linkage (black circles), average linkage (red circles), and complete linkage (blue circles). In all three cases we observe a bell shaped curve of the ARI as a function of the threshold. Therefore in all three cases a single maximum exists for a specific value of the threshold. In the figure we show the values of the ARI computed by varying the threshold in steps of 0.1. However in the proximity of the maximum we have used an increment of the threshold variation set to 0.01. With this resolution we have verified that the highest value of the ARI are observed for threshold values equal to 0.80, 1.09 and 1.31 for the single, average and complete hierarchical tree respectively. The maximal values of the ARI observed are 0.806, 0.923, and 0.795 for single, average, and complete respectively. These are quite high values and therefore for the optimal threshold values the partitions of the hierarchical trees are pretty similar to the partitions obtained from Bonferroni statistical validation and the successive community detection done with Infomap. The highest similarity is observed for the average linkage hierarchical clustering. 
\begin{center}
\begin{figure}
\includegraphics[scale=0.49]{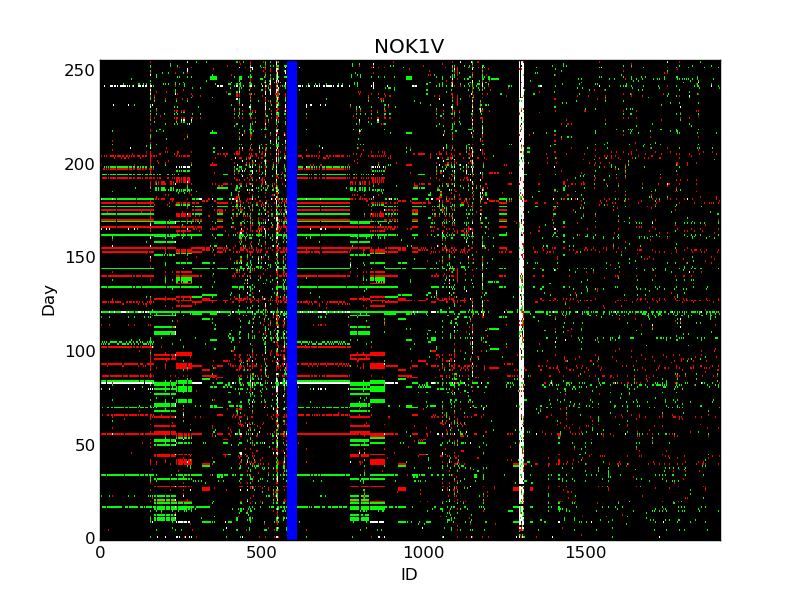}
\caption{\scriptsize{Color code representation of the trading profile of investors trading the Nokia. The left part of the figure (limited by a blu vertical bar) shows trading profiles of the clusters of 576 Nokia investors  in the Bonferroni statistically validated network (same plot as in the left panel of Fig. \ref{F5}) while the right part are the clusters obtained from the hierarchical tree of the average clustering by using the 1.09 threshold. The overlapping clusters are ordered with the ordering of Fig. \ref{F5}.}}
\label{F7}
\end{figure}
\end{center}

To provide a visual comparison of the similarity of the two partitions, in Fig. \ref{F7} we show the color code representation of the trading profile of investors trading the Nokia stocks grouped both as defined by the Bonferroni statistically validated network and as defined by the average clustering hierarchical tree when a threshold obtained by maximizing the ARI is used. In the specific case the threshold value is set to 1.09 and the ARI is 0.923.
The Figure shows the high overlap quantified by the high value of the ARI index. It also shows that the hierarchical tree approach extends the information available to a large number of investors. In fact when a threshold of value 1.09 is used 1419 Nokia investors are detected in 319 clusters of at least 2 investors. The number of investors of the hierarchical tree partition is significantly larger than the number of 576 Nokia investors observed by the Bonferroni statistically validated network. The role of the type of the hierarchical clustering used is not crucial for the largest clusters although we have empirical evidence that the highest values of ARI are typically observed for the average clustering.  The use of the single linkage is increasing the probability of observing large clusters whereas the complete linkage select largest clusters or relatively smaller size. The average linkage provide an intermediate behavior.
\begin{center}
\begin{figure}
\includegraphics[scale=0.45]{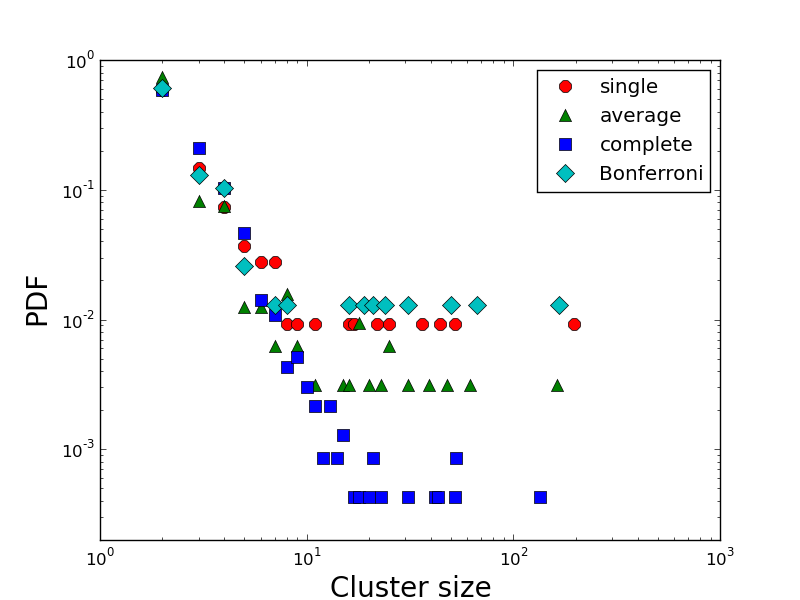}
\caption{\scriptsize{Probability density function of the cluster size of the partitions obtained with the Bonferroni statistically validated network partitioned by the Infomap algorithm  (light blue diamond), single linkage (red circle), average linkage (green triangle), and complete linkage (blue square). The stock is Nokia.}}
\label{F8}
\end{figure}
\end{center}

In Fig. \ref{F8} we show the probability density function of the size (in number of investors) of clusters observed for the Bonferroni partition, and for the three partitions obtained with the hierarchical clustering algorithms. All the three hierarchical clustering methods reproduce well the behavior observed for the Bonferroni partition. The number of partitions and the number of investors present in the partitions are higher for the complete linkage, intermediate for the average linkage and lower for the single linkage. By considering the value of the ARI for the investors which are present both in the Bonferroni partition and in the complete linkage partition, the broader covering of investors and partitions of the complete is probably obtained at a cost of a lower precision of the partitioning. In the present system and in the present example the average linkage provides the best choice for partitioning a number of investors larger than the Bonferroni partition which is maintaining approximately the same precision as the Bonferroni for the investors selected with the statistically validated approach.   

The observation that the Bonferroni statistically validated networks is providing highly precise but not necessarily accurate clusters is not surprising. In fact it is known that the Bonferroni correction is too restrictive. This choice ensures a high level of precision (because the number of false positive is minimal) but on the other hand it might be associated with a high number false negatives and therefore be characterized by a relatively low level of accuracy. The cluster detection based on the correlation measure is therefore providing results which might be a bit less precise but more accurate. The problem with the correlation approach is that the thresholding process is not supported by any theoretical indication. 

Our observation of significant overlap of the clusters obtained with the two distinct approaches suggests the effectiveness of a new methodology using Bonferroni statistically validated network partitioned with an efficient community detection algorithm together with a hierarchical clustering procedure. Within this approach the optimal threshold to be used in the hierarchical clustering procedure can be determined by using the precise information obtained with the Bonferroni approach. Finally, when the threshold is determined, one obtains the partition from the hierarchical tree. This last partition covers a larger number of investors which is suggesting a higher level of accuracy and perhaps a bit lower level of precision.

\section{Discussion and conclusions}
\label{Con}

We have verified that the number of distinct investors participating to the process of price discovery in a regional financial market can be rather limited for some of the traded stocks. In Table \ref{Tab1} we note that the number of Finnish investors trading the Nokia stock was 41729 and that 7824 of them (see Table \ref{Tab2}) were performing at least five transaction during 2003. These numbers can be seen as reasonably large for the process of price discovery (especially by considering that a lot of other transactions are done by foreign investors acting as nominee registered). However, similar numbers are not observed for the other stocks comprising the OMXH25 index. In fact, in addition to Nokia only other six stocks have more than 10,000 investors trading the stock at least once in 2003 (they are ELIAV, FUM1V, NDA1V, OUT1V, TLS1V, and UPM1V see Table \ref{Tab1}) but only STERV and UPM1V have more than 1,000 investors performing at least five transactions during 2003. The system is therefore highly heterogeneous in terms of transaction and in terms of number of investors investing in a given stock.

We have verified that this heterogeneity does not prevent the use of a correlation-based approach performed with categorical variables. In fact we have shown that it is possible to device a method based on categorical variables of states buy, sell and buy/sell and obtain and analyze a dissimilarity matrix computed by using a Jaccard measure between the vectors describing the trading profile of a pair of investors. The method is able to reveal the hierarchical structure of the trading profiles of heterogeneous investors. We show that our correlation based method detects hierarchical structures which are overlapping with the cluster structures obtained with the approach of statistically validated networks when an appropriate threshold of the hierarchical trees is used. 

We also propose a way to combine the two methods to extract additional information from the data. In fact the method based on statistically validated network and community detection is giving results of high precision but of unknown accuracy. The combination of the approach of statistically validated networks with the correlation-based approach expands the set of information available about the clusters of investors suggesting an increase of the accuracy at a minimal cost of decrease in terms of precision (i.e. in terms of increase of false positive). We propose to use a combination of the two methods when there are indications that the accuracy of the statistical validation network approach is too limited to properly describe the system of interest.   





\end{document}